\documentclass[prl, twocolumn]{revtex4-1}

\usepackage{graphicx}
\usepackage{color}
\usepackage{verbatim}
\usepackage{subfigure}
\usepackage{amssymb}
\usepackage{amsmath}

\usepackage{hyperref}

\begin{document}

\title{Generalized hydrodynamics of active polar suspensions}

\author{Dibyendu Mandal,$^{1}$ Katherine Klymko,$^{2}$ and Kranthi K. Mandadapu$^{3, 4}$}
\affiliation{
$^1$Department of Physics, University of California, Berkeley, CA 94720, U.S.A.\\
$^2$Department of Chemistry, University of California, Berkeley, CA 94720, U.S.A.\\
$^3$Department of Chemical and Biomolecular Engineering, University of California, Berkeley, CA 94720, U.S.A. \\
$^4$Chemical Sciences Division, Lawrence Berkeley National Laboratory, Berkeley, CA 94720, U.S.A.}

\date{\today}

\begin{abstract} 
We utilize a generalized Irving-Kirkwood procedure to derive the hydrodynamic equations of an active matter suspension with internal structure and driven by internal torque. 
The internal structure and torque of the active Brownian particles give rise to a balance law for internal angular momentum density, making the hydrodynamic description a polar theory of continuum mechanics.
We derive exact microscopic expressions for the stress tensor, couple stress tensor, internal energy density, and heat flux vector. 
Unlike passive matter, the symmetry of the stress tensor is broken explicitly due to active internal torque and the antisymmetric component drives the internal angular momentum density.
These results provide a molecular basis to understand the transport characteristics and collectively provide a strategy to develop the theory of linear irreversible thermodynamics of active matter.
\end{abstract}

\maketitle


Systems composed of self-activating units are called active matter~\cite{Ramaswamy_2010, Romanczuk_2012, Marchetti_2013, Yeomans_2014, Menzel_2015, Bechinger_2016}~\footnote{Whether a system is active or simply driven depends on the definition of the system. A system looks active if we do not look at internal mechanisms of the system and consider it as a whole. For example, in myxobacteria the internal organs provide a driving mechanism but the bacteria as a whole look active. Similarly, if we consider a colloid that is driven by an external magnetic field the colloid may look active even though it is actually driven by the magnetic force.}. 
Active matter systems exhibit interesting emergent behavior such as enhanced tracer diffusion~\cite{Wu_2000, Leptos_2009, Mino_2011}, giant number fluctuations~\cite{Narayan_2007}, motility induced phase separation (MIPS)~\cite{Tailleur_2008, Fily_2012, Buttinoni_2013, Redner_2013, Stenhammer_2013, Mognetti_2013, Stenhammer_2015, Cates_2015, Redner_2016}, spontaneous emergence of rectification~\cite{Wan_2008, Angelani_2009, DiLeonardo_2010, Angelani_2010, Ghosh_2013, Kaiser_2014, Mallory_2014_PRE_I}, collective motion~\cite{Vicsek_1995, Toner_1998, Toner_2005, Vicsek_2012, Cavagna_2014}, pattern formation~\cite{Guillaume_2010, Zuiden_2016}, odd viscosity~\cite{Bannerjee_2017}, and topological localization~\cite{Dasbiswas_2017}.
Active matter systems are abundant in the biological domain, ranging from the cytoplasmic fluid and bacterial colonies to flocks of birds and herds of animals~\cite{Vicsek_1995, Toner_1998, Toner_2005, Howse_2007, Julicher_2007, Narayan_2007, Cates_2012, Vicsek_2012, Cavagna_2014}. 
On the technological side, they provide a viable candidate for the development of colloidal programmable matter~\cite{Palacci_2013, Ni_2015, Goodrich_2016, Mano_2017}. 

          
Active matter systems provide an opportunity to revisit the notions of statistical mechanics and condensed matter physics from a fresh nonequilibrium perspective. 
The concept of pressure has recently received particular emphasis~\cite{Mallory_2014_PRE_II, Takatori_2014_PRL, Solon_2015, Solon_2015_PRL, Winkler_2015, Speck_2016_PRE, Nikola_2016, Joyeux_2016, Marconi_2016_arXiv, Joyeux_2017, Marconi_2017_arXiv, Fily_2017, Sandford_2017_arXiv}.
For example, (osmotic) pressure of an active matter suspensions of spherical, self-propelled particles (the so-called active Brownian particles, ABPs) has been suggested to be a \emph{nonequilibrium} state function in absence of internal torques~\cite{Solon_2015}. 
The justification of this suggestion comes from the observation that pressure is independent of the interaction of the ABPs with the walls of the suspension. 
Alternatively, in the context of MIPS, Cahn-Hilliard type arguments have been used to derive the chemical potential and equivalently the pressure~\cite{Speck_2014, Takatori_2015_PRE}. 
To fully specify the dynamics and transport characteristics of active matter, however, we need to specify not just the pressure but also the stress and couple stress tensors~\cite{Ericksen_1960, Ericksen_1961, Dahler_1961, Dahler_1963, Leslie_1966, Leslie_1968, Evans_1976, deGroot_1984, Stokes_1966, Stokes_1984, Travis_1997, Spencer_2004, Tsai_2005, Stark_2005, Lau_2009}.
The purpose of this Letter is to introduce these concepts on a molecular basis. 
 
\begin{figure}
\centering
\includegraphics[width= 0.37 \textwidth]{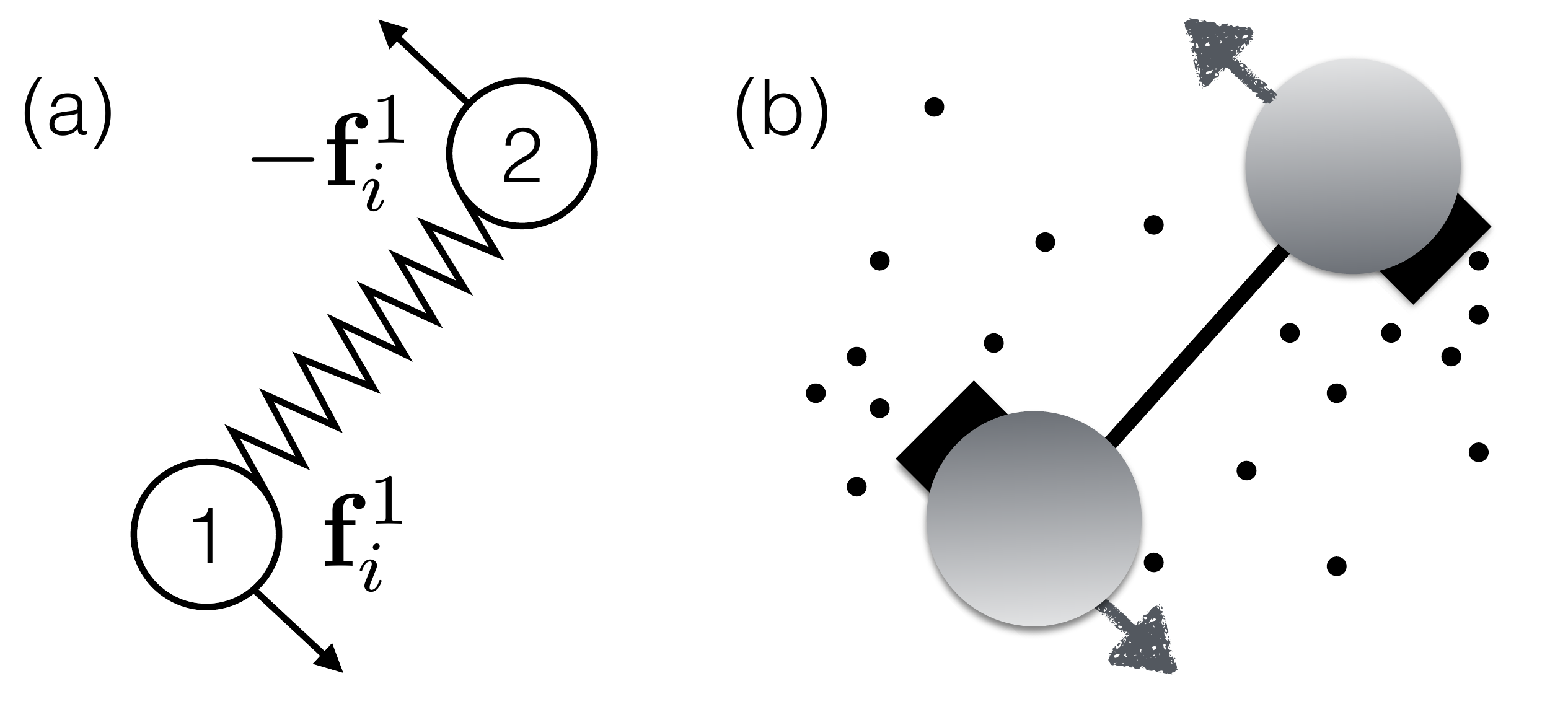}
\caption{(a) Cartoon of an individual ADP. The potential between the two particles of an ADP is depicted by a linear spring. An ADP is driven by an internal torque arising from forces $\mathbf{f}_1$ and $\mathbf{f}_2 = - \mathbf{f}_1$. (b) The ADPs can be realized by tethering together two colloidal spheres of Ref.~\cite{Palacci_2013} with their (black) hematite protrusions in opposite directions.}
\label{fig:dumbbell}
\end{figure}

We utilize a generalized Irving-Kirkwood procedure to derive the balance laws of mass, linear momentum, angular momentum, moment of inertia, total energy, and internal energy governing the macroscopic behavior of active fluids starting from their microscopic dynamics~\cite{Irving_1950, Grad_1952, Noll_1955_Original, Noll_1955, Dahler_1959, Evans_2008, Mandadapu_2012}. 
Using this procedure, we derive exact microscopic expressions for the stress and couple stress tensors (and hence also pressure, which is just the negative trace of the stress tensor)~\footnote{Our approach must be contrasted with the phenomenological approach where the hydrodynamic fields are simply postulated rather than derived based on microscopic dynamics~\cite{Foffano_2012, Furthauer_2012_EPJE, Furthauer_2012_NJP, Prost_2015}. Also, we do not assume the existence of any effective Hamiltonian~\cite{Marconi_2015_SM} or free energy in these far-from-equilibrium systems~\cite{Lau_2009, Ramaswamy_2010, Marchetti_2013, Thampi_2014, Thampi_2015, Tiribocchi_2015, Nardini_2016}.}. 
The stress tensor can be useful to calculate the surface tension of phase-separated interfaces~\cite{tenWolde_1998} and rheological properties. 
We also derive an exact microscopic expression for the heat flux vector, thus paving the way for a future study on the linear irreversible thermodynamics of active systems in terms of internal entropy production. 
Stress tensors and heat flux vectors with microscopic expressions can be used in the Green-Kubo formulae~\cite{Kubo_1957} to calculate the transport coefficients of viscosity and thermal conductivity~\cite{Mandadapu_2010, Jones_2012}.
Our model system consists of active dumbbell particles (ADPs) with internal torque (Fig.~\ref{fig:dumbbell})~\footnote{We have considered underdamped dynamics, not just the overdamped limit~\cite{Winkler_2015, Winkler_2016}. Also, we have considered internal torque, not self propulsion as in Refs.~\cite{Joyeux_2016, Joyeux_2017}. Finally, ADPs are assumed to be deformable and not rigid~\cite{Nguyen_2014}.}. 
The passive version of the ADP is a classic model for polymer dynamics and rheology~\cite{Bird_1987}. 
The original Irving-Kirkwood procedure was developed for passive, point particles, without any internal rotational degree of freedom~\cite{Irving_1950}. 
Our work can be considered to be a generalization of the procedure to structured (\emph{polar}) fluids. 
Furthermore, we have discovered the existence of asymmetric stress tensor for ADP suspensions, which is absent in passive and even self-propelled active Brownian suspensions (such as ABP suspensions). 
We have also suggested a potential experimental realization of our results.


An ADP colloidal ``molecule" $i$ is composed of two ``atoms" $(i, 1)$ and $(i, 2)$, as shown in Fig.~\ref{fig:dumbbell}(a).  
The mass, position, and momentum of the atom $(i, \alpha)$ are denoted by $m_i^\alpha$, $\mathbf{x}^{\alpha}_i$, and $\mathbf{p}^{\alpha}_i$, respectively. 
The interaction force on $(i, \alpha)$ from $(j, \beta)$ is denoted by $\mathbf{F}_{ij}^{\alpha \beta} = - \partial u_{ij}(\mathbf{x}^{\alpha}_i,\mathbf{x}^{\beta}_j) / \partial \mathbf{x}^{\alpha}_{i}$ for some two-body potential energy $u_{ij}(\mathbf{x}^{\alpha}_i,\mathbf{x}^{\beta}_j)$~\footnote{In general, we have $u_{i\neq j} \neq u_{ii}$. In the companion paper we use a more detailed notation with $u_{ii} = u_\text{s}$ and $u_{i \neq j} = u_2$}. 
We assume $u_{ij}(\mathbf{x}^{\alpha}_i,\mathbf{x}^{\beta}_j) = u_{ij}(|\mathbf{x}^{\alpha}_i - \mathbf{x}^{\beta}_j|)$.  
The ADPs are assumed to be suspended in a stationary solution. 
Due to their asymmetric interactions with the surrounding fluid, the ADPs are subjected to an internal torque as shown in Fig.~\ref{fig:dumbbell}(a) with the active forces $\mathbf{f}^1_i = - \mathbf{f}^2_i$.
We assume that the active forces are always perpendicular to the line joining the two atoms of an ADP molecule~\footnote{This restriction can be relaxed to have a general model of ADPs that includes many other models of active Brownian particles, for example the so-called active Brownian particles (ABPs)~\cite{Redner_2013}.}. 
A possible realization of ADPs involves the setup of Ref.~\cite{Palacci_2013} where a colloidal sphere has a protruding hematite cube that can catalyze the decomposition of hydrogen peroxide in the presence of blue light. 
An ADP can be created by tethering together two such colloidal spheres with their hematite protrusions in opposite directions, as shown in Fig.~\ref{fig:dumbbell}(b).  
Because the ADPs are inside a solution, they feel the thermal Langevin forces as well, a drag force $- \zeta \mathbf{p}^{\alpha} _i / m_i^\alpha$, for drag coefficient $\zeta$, and a Gaussian white noise force, $\sqrt{2 k_\text{B} T } \xi_i^\alpha$, with $\langle \xi_i^\alpha \rangle = 0$ and $\langle \xi_i^\alpha(t) \xi_j^\beta(t')  \rangle = \delta_{ij} \delta_{\alpha \beta} \delta(t - t')$ for Kronecker deltas $\delta_{ij}$ and $\delta_{\alpha \beta}$ and Dirac delta function $\delta(t - t')$. 
Here $k_\text{B}$ is the Boltzmann constant and $T$ is the temperature of the solution. 
The equations of motion are
\begin{equation}
\label{eq:EqnMot} 
\dot{\mathbf{x}}^{\alpha}_i = \mathbf{p}^{\alpha}_i/m_i^\alpha, \quad \dot{\mathbf{p}}^{\alpha}_i = \sum_{j \beta} \mathbf{F}^{\alpha \beta}_{ij} + \mathbf{f}^{\alpha}_i + \mathbf{f}_{i, \text{Th}}^\alpha,
\end{equation}
with $\mathbf{f}_{i, \text{Th}}^\alpha =  - \zeta \mathbf{p}^{\alpha} _i / m_i^\alpha +  \sqrt{2 k_\text{B} T \zeta} \xi_i^\alpha$. 
Here and in what follows, $\dot{( * )} = \mathrm{d} (*) / \mathrm{d} t$ denotes the total time derivative. 
In the current setup, we have assumed that all the body force comes from the solvent. 
The situation can be generalized in a straightforward way to include other body forces, such as self-propulsive (convective) forces (as in ABPs~\cite{Redner_2013}) and gravity.


\begin{figure}[t]
\centering
\includegraphics[width=0.4\textwidth]{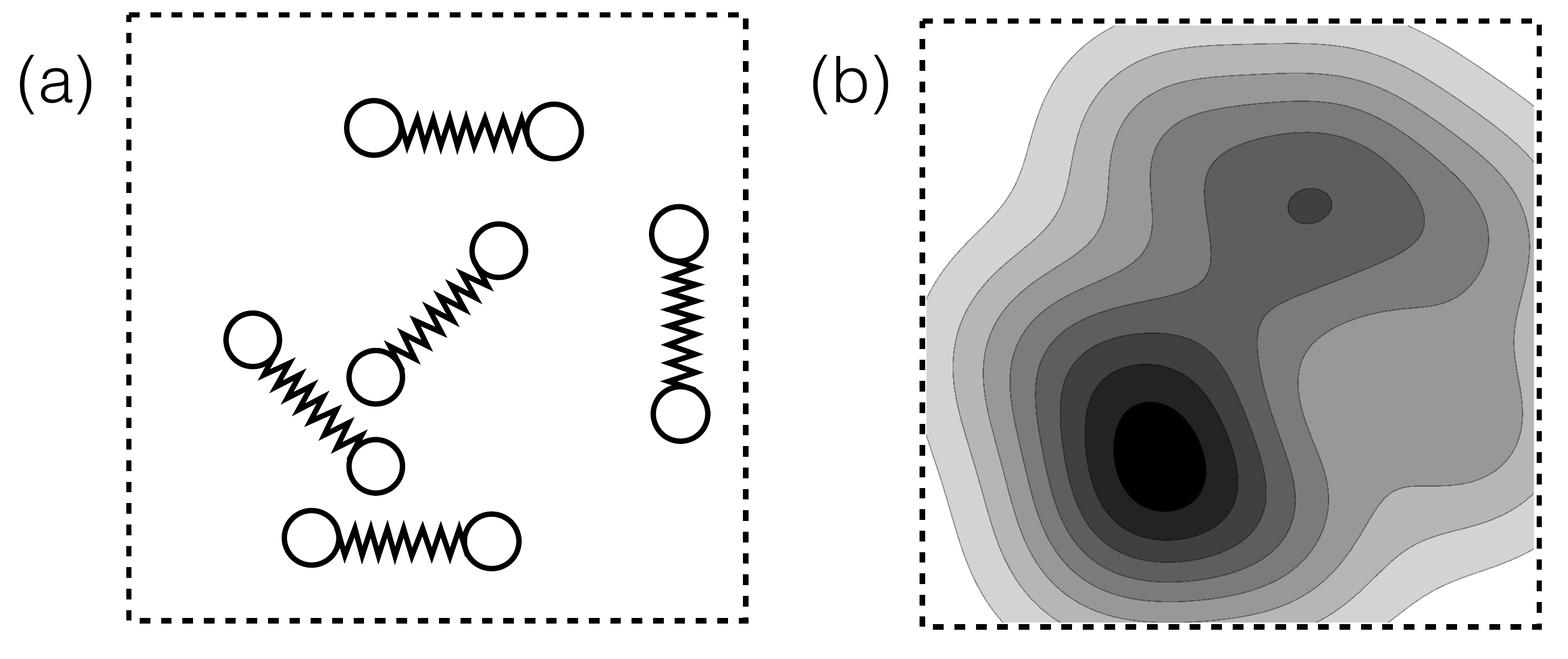}
\caption{Illustration of the coarse-graining function. The particle picture in (a) is replaced by a continuum picture in (b) by the coarse graining procedure described in the text. We have assumed a Gaussian form for the coarse-graining function $\Delta_i^\alpha$, about $\mathbf{x}_i^\alpha$ in this picture. The bottom-left corner is darker because there are more particles in that corner. There is an intermediate asymptotic length-scale associated with the coarse-graining function such that the final results are insensitive with respect to the size of the corse-graining function within its range~\cite{Ulz_2013}.} 
\label{fig:CG}
\end{figure}

A key element in our discussions is a \textit{coarse-graining function} $\Delta(\mathbf{x} - \mathbf{x}_i^\alpha) \equiv \Delta_i^\alpha$ which is a unimodal function concentrated about $\mathbf{x}_i^\alpha$, and whose effect is to replace the particle $(i, \alpha)$ by a smeared density distribution with equal total mass~\cite{Mandadapu_2012}. 
This helps us make a transition from the microscopic particulate picture to a continuous hydrodynamic picture, as illustrated in Fig.~\ref{fig:CG}.
This is different from considering an average over several repetitions of the dynamics with different initial conditions and noise realizations. 
Fluctuations from the initial conditions and noise realizations are still present in our picture, albeit in a smeared and less pronounced manner~\footnote{This contrasts with other approaches where these ensemble averages are always performed~\cite{Baskaran_2009, Steffenoni_2016_arXiv_I, Fily_2017, Gao_2017, Wittkowski_2017}.}.
In the following, we do not consider any specific form for $\Delta_i^\alpha$, other than the following property, $\partial \Delta^{\alpha}_i / \partial \mathbf{x}^{\alpha}_i = -\partial \Delta^{\alpha}_i / \partial \mathbf{x}$, which implies $\Delta_i^\alpha$ is a function of just the difference $\mathbf{x} - \mathbf{x}_i^\alpha$~\footnote{Textbook discussions usually consider the Dirac delta function for $\Delta(\mathbf{x} - \mathbf{x}_i^\alpha)$. The lengthscale associated with $\Delta(\mathbf{x} - \mathbf{x}_i^\alpha)$ should be larger than the size of the particles but smaller the legthscale of hydrodynamic fluctuations. A lower bound was proposed in Ref.~\cite{Ulz_2013} based on the correlation length of interparticle potential energies.}.
Based on the coarse graining function $\Delta_i^\alpha$, we can introduce local densities for mass ($\rho$), momentum ($\rho \mathbf{v}$), angular momentum ($\rho \mathbf{L}$), and energy ($\rho e$): 
\begin{equation}
\label{eq:Density}
\rho \{ 1, \mathbf{v}, \mathbf{L}, e \} = \sum_{i \alpha} \{m^{\alpha}_i, \mathbf{p}^{\alpha}_i, \mathbf{x}_i^\alpha \times  \mathbf{p}^{\alpha}_i, e_i^\alpha \} \Delta_i^\alpha
\end{equation}
with energy per particle $ e_i^\alpha =  (1/2) \sum_{i \alpha} \big[ m_i^\alpha (v_i^\alpha)^2  + \sum_{j \beta} u_2(\mathbf{x}^{\alpha}_i, \mathbf{x}^{\beta}_j) \big]$~\footnote{We have distributed the potential energies equally between the corresponding pair of atoms.}.


We now use the expressions in Eq.~\eqref{eq:Density} to derive the balance equations for local densities. 
We assume that the point of observation is moving with the \emph{barycentric} velocity $\mathbf{v}(\mathbf{x}, t)$, introduced in Eq.~\eqref{eq:Density} (Lagrangian point of view).
This is equivalent to assuming that $\mathbf{x}$ itself is a function of time with $\dot{\mathbf{x}} = \mathbf{v}(\mathbf{x}(t), t)$ . 
The method to derive balance laws involves (i) differentiation of the densities defined in Eq.~\eqref{eq:Density} with respect to time; (ii) usage of the equations of motion (Eq.~\eqref{eq:EqnMot}), the property $\partial \Delta^{\alpha}_i / \partial \mathbf{x}^{\alpha}_i = -\partial \Delta^{\alpha}_i / \partial \mathbf{x}$, and $\dot{\mathbf{x}} = \mathbf{v}(\mathbf{x}(t), t)$; and (iii)  utilization of tensor identities to express the resulting equations in forms considered in continuum mechanics~\cite{deGroot_1984, Stokes_1984, Spencer_2004}, i.e., in terms of surface (divergence) and volumetric terms. 
The details are presented in~\footnote{K. Klymko, D. Mandal, and K. K. Mandadapu, \emph{in preparation}.}.


\begin{figure}[!t]
\centering
\includegraphics[width=0.4\textwidth]{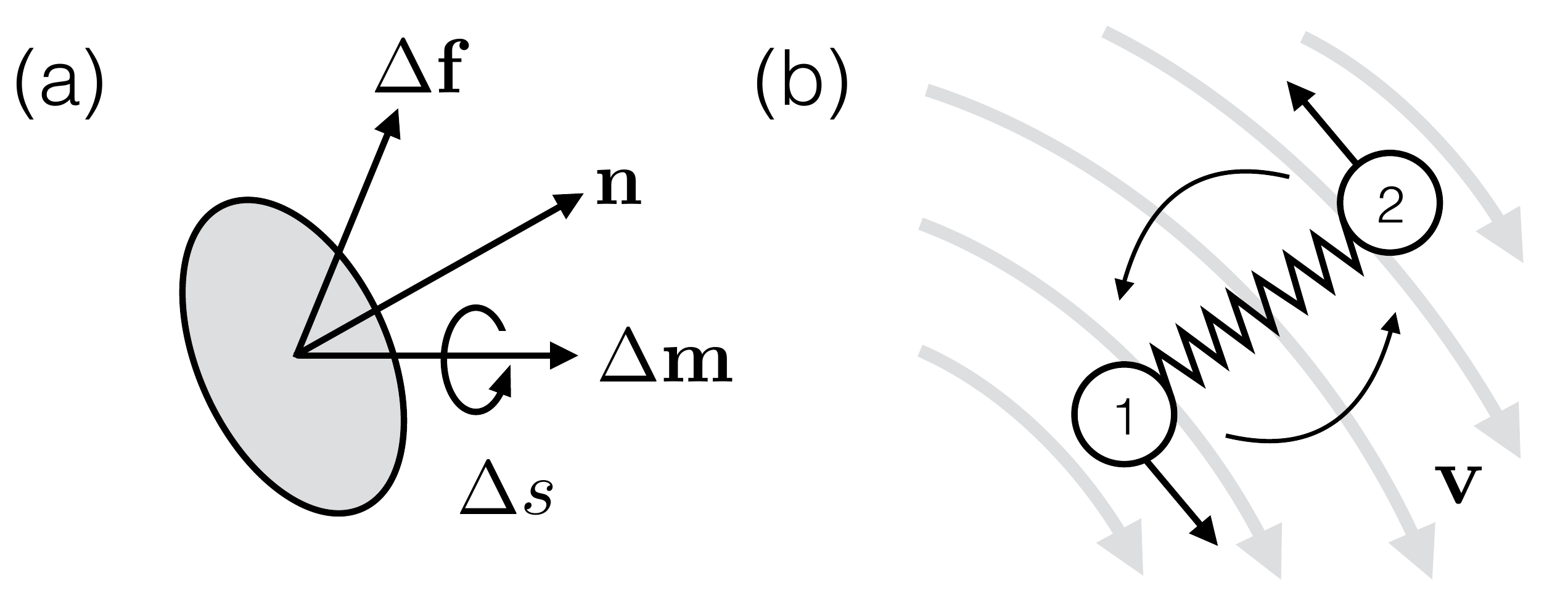}	
\caption{(a) Geometric definitions of stress and couple stress tensors. For any infinitesimal surface element with area $\Delta s$ and normal vector $\mathbf{n}$, an infinitesimal force $\Delta \mathbf{f}$ and an infinitesimal moment $\Delta \mathbf{m}$ can be applied. Then, the stress tensor $\mathbf{T}$ and the couple stress tensor $\mathbf{C}$ are related to the forces and moments by $\lim_{\Delta s \rightarrow 0} \Delta \mathbf{f} / \Delta s = \mathbf{T} \cdot \mathbf{n}$ and  $\lim_{\Delta s \rightarrow 0} \Delta \mathbf{m} / \Delta s = \mathbf{C} \cdot \mathbf{n}$, respectively~\cite{Stokes_1984}. Here, the dot product between a tensor and vector is equivalent to matrix multiplication. (b) An ADP can rotate about an instantaneous axis because of its internal structure and torque. This internal spin leads to an internal angular momentum.}
\label{fig:Couples}
\end{figure}

For mass balance, we get the usual continuity equation, $\dot{\rho} + \rho  \boldsymbol{\nabla} \cdot \bold{v} = 0$.
The linear momentum balance can be derived to be of the form
\begin{equation}
\label{eq:MomBal}
\rho\dot{\mathbf{v}} = \boldsymbol{\nabla} \cdot \mathbf{T} +\rho\mathbf{b},
\end{equation}
where $\mathbf{T}$ is the stress tensor and $\mathbf{b}$ is the body force per unit mass. 
A pictorial definition of $\mathbf{T}$ based on continuum mechanics perspective is given in Fig.~\ref{fig:Couples}(a). 
Pressure is given by $p = - \text{Tr}({\mathbf{T}})$. 
This mechanical notion of pressure is well defined in all fluids, equilibrium or not.
For fluids in thermodynamic equilibrium, the mechanical pressure coincides with the statistical mechanical definition of pressure derived from the partition function formalism~\cite{McQuarrie_1976, Chandler_1987}. 
Exact microscopic expressions derived for $\mathbf{T}$ and $\mathbf{b}$ are given in Table~\ref{tab:Exact} (second column), where we see that stress tensor $\mathbf{T}$ is composed of three components -- kinetic $\mathbf{T}^\text{K}$, potential $\mathbf{T}^\text{P}$, and active $\mathbf{T}^\text{A}$~\footnote{In the accompanying paper, we have $\mathbf{T}^\text{P} = \mathbf{T}^\text{V} + \mathbf{T}^\text{S}$. The symbol $\otimes$ in these expressions indicates the dyadic product, i.e., the products $\mathbf{a} \otimes \mathbf{b}$ of any two vectors $\mathbf{a}$ and $\mathbf{b}$ is a second order tensor with elements $a_i b_j$. Note that, in general, the dyadic product does not yield a symmetric tensor, i.e., $\mathbf{a} \otimes \mathbf{b} \neq \mathbf{b} \otimes \mathbf{a}$.}. 
The kinetic part of the stress tensor $\mathbf{T}^\text{K}$ includes contributions from the fluctuations of the velocity of particles with respect to the barycentric velocity $\mathbf{v}$~\footnote{This is not the thermal part of the velocity, which we define after Eq.~\eqref{eq:EngDec}.}. 
The potential part $\mathbf{T}^\text{P}$ contains contributions from the interactions between the particles from both intermolecular and harmonic spring interactions. 
The $\mathbf{T}^\text{K}$ and $\mathbf{T}^\text{V}$ parts of the stress tensor are common to any fluids as was originally derived by Irving and Kirkwood. 
However, as can be seen in Table~\ref{tab:Exact}, there exists a new contribution $\mathbf{T}^\text{A}$ to the stress tensor that comes from the active forces on the dumbbell. 
This contribution is specific to active matter and differentiates the system from passive matter. 
In $\mathbf{T}^\text{P}$ and $\mathbf{T}^\text{A}$, we needed to introduce the Noll bond function $b_{ij}^{\alpha \beta} = \int_0^1 \mathrm{d} \lambda \ \Delta (\mathbf{x} - \lambda \mathbf{x}^{\alpha}_i +\mathbf{x}^{\alpha \beta}_{ij})$ with $\mathbf{x}_{ij}^{\alpha \beta} = \mathbf{x}_i^\alpha - \mathbf{x}_j^\beta$~\cite{Noll_1955_Original, Noll_1955, Evans_2008}. 
While $\mathbf{T}^\text{K}$ and $\mathbf{T}^\text{P}$ are symmetric tensors, $\mathbf{T}^\text{A}$ is not. 
This asymmetry in the stress tensor arises due to the active couples from the active forces $\mathbf{f}_i^\alpha$, which break the microscopic rotational symmetry at the level of particle dynamics~\footnote{In other words, the dynamics of the ADPs is nonHamiltonian in nature~\cite{Mandal_2016_PRE, Mandal_2017_arXiv}}.
The existence of asymmetric stress tensor in ADP suspensions is one of the central findings of this Letter. 
This asymmetry is absent in passive and active self-propelled (convective) suspensions~\footnote{We assume only pairwise radial interactions.}. 

\begin{table*}
\begin{tabular}{lll}
Quantity & \qquad Linear momentum & \qquad Angular momentum \\
\hline
Stress tensor & \qquad $\mathbf{T} = \mathbf{T}^\text{K} + \mathbf{T}^\text{P} + \mathbf{T}^\text{P}$ & $\qquad \bold{C} =  \bold{C}^\text{K} +  \bold{C}^{\text{P}}+  \bold{C}^{\text{A}}$ \\
\quad (i) Kinetic & \qquad $\bold{T}^\text{K} = - \sum_{i \alpha} m^{\alpha}_i \left( \mathbf{v}_i^\alpha - \mathbf{v} \right) \otimes \left( \mathbf{v}_i^\alpha - \mathbf{v} \right) \Delta_i^\alpha$ & \qquad $\bold{C}^\text{K} = - \sum_{i \alpha} \left[ (\bold{x}^{\alpha}_i - \bold{x}) \times \bold{p}^{\alpha}_i \right] \otimes \left( \mathbf{v}_i^\alpha - \bold{v} \right) \Delta_i^\alpha$ \\
\quad (ii) Potential & \qquad $\bold{T}^\text{P} = - \frac{1}{2} \sum_{i j \alpha \beta} \mathbf{F}^{\alpha \beta}_{ij} \otimes  \mathbf{x}^{\alpha \beta}_{ij} b^{\alpha \beta}_{ij}$ & \qquad $\bold{C}^{\text{P}} = - \frac{1}{2} \sum_{i j \alpha \beta} \left[ (\bold{x}^{\alpha}_i -\bold{x}) \times \bold{F}^{\alpha \beta}_{ij} \right] \otimes \bold{x}^{\alpha \beta}_{ij} b^{\alpha \beta}_{ij}$\\
\quad (iii) Active & \qquad $\mathbf{T}^\text{A} = - \sum_i \mathbf{f}_i^1 \otimes \mathbf{x}^{12}_{ii} b^{12}_{ii}$ & \qquad $\bold{C}^{\text{A}} = - \sum_i \left[ (\bold{x}^1_i -\bold{x}) \times \bold{f}_i^1 \right] \otimes \bold{x}^{12}_{ii} b^{12}_{ii}$\\
Body vector & \qquad $\rho \mathbf{b} = \sum_{i, \alpha} \mathbf{f}_{i, \text{Th}}^\alpha \Delta_i^\alpha$ & \qquad $\rho \bold{G} = \sum_{i} \big[\bold{x}^{12}_{ii} \times \bold{f}_i^1 \Delta_i^2 + \sum_\alpha \left( \bold{x}^{\alpha}_i -\bold{x} \right) \times \mathbf{f}_{i, \text{Th}}^\alpha \Delta_i^\alpha \big]$
\end{tabular}
\caption{Exact microscopic expressions of stress and couple stress tensors. Also included are the body forces and body couples. The second column gives the results for the balance of linear momentum and the third column gives the results for angular momentum. The principal benefit of such a result is that we can theoretically derive the stress and couple stress tensors, and body forces and couples in the hydrodynamic equations of active fluids, exactly, without making any assumption. }
\label{tab:Exact}
\end{table*}


The balance equation for total angular momentum can be derived to be of the form
\begin{equation}
\label{eq:AngMomBal}
\rho \dot{\bold{L}} = \boldsymbol{\nabla} \cdot  \bold{C}  + \rho \bold{G} + \bold{A}+ \bold{x} \times \rho \bold{b} + \bold{x} \times \left( \boldsymbol{\nabla} \cdot \bold{T}\right), 
\end{equation}
where $\textbf{C}$ is the total couple stress tensor, $\textbf{G}$ is the body couple per unit mass, and $\mathbf{A}$ is the antisymmetric component of the stress tensor, $A_i = \epsilon_{ijk} T_{jk}$, where $\epsilon_{ijk}$ is the Levi-Civita symbol. 
A geometric definition of $\textbf{C}$, from a purely continuum mechanics point of view, is illustrated in Fig.~\ref{fig:Couples}(a). 
Exact microscopic expressions for $\textbf{C}$ and $\textbf{G}$ are given in Table~\ref{tab:Exact} (third column). 
As with the stress tensor $\mathbf{T}$, the couple stress tensor $\mathbf{C}$ is composed of three components -- kinetic $\mathbf{C}^\text{K}$, potential $\mathbf{C}^\text{P}$, and active $\mathbf{C}^\text{A}$ -- with similar origins~\footnote{In the accompanying paper, we have $\mathbf{C}^\text{P} = \mathbf{C}^\text{V} + \mathbf{C}^\text{S}$.}. 
As can be seen from the expressions in Table~\ref{tab:Exact}, all the components of $\textbf{C}$ are given by the moments of the forces acting on the atoms with respect to the center $\mathbf{x}$ of the coarse-graining volume. 
These terms are analogous to the terms in the stress tensor $\mathbf{T}$, with moments replacing the forces. 
The microscopic derivation of the balance of angular momentum was not considered by Irving and Kirkwood.
Our work can be seen as a generalization of their work to structured particles with internal torques. 
The couple stress tensor $\mathbf{C}$ for passive matter is often negligible because the relaxation time of the couple stresses is generally small compared to the observation timescale~\cite{deGroot_1984}.
Therefore, the change of angular momentum is ignored leading to the conclusion that the stress tensor is symmetric~\footnote{Note that systems modeled by three-body potentials, such as Stillinger-Weber potentials [PhD Thesis, K. K. Mandadapu] can contribute three-body terms to the stress tensor, which break the symmetry. Such terms can affect the balance of angular momentum through the term $\mathbf{A}$.}. 
In the case of ADPs, $\mathbf{C}^A$ results from the internal torque and is not negligible.
In this case the angular momentum effects cannot be ignored, the stress tensor is not symmetric, and there is a possibility for coupled linear and angular momentum phenomena~\cite{Stark_2005, Zuiden_2016}. 
The body couple $\textbf{G}$ comes from the moments of body forces $\mathbf{f}_{i, \text{Th}}^\alpha$ and $\mathbf{f}_i^\alpha$ with respect to $\mathbf{x}$. 
It is interesting to see that the torque from the active forces $\mathbf{f}_i^\alpha$ contributes to both the surface term $\textbf{C}$ and the body term $\textbf{G}$. 
The last two terms on the right of Eq.~\eqref{eq:AngMomBal} denote the moments induced by the body forces and the surface forces with respect to the origin in the laboratory frame~\footnote{Note that the divergence of a second order tensor is defined as  $(\boldsymbol{\nabla} \cdot \bold{S})_i = \sum_j (\partial / \partial x_j) S_{ij}$.}.


The ADPs can have internal rotations or spin because of their internal structure and torque as illustrated in Fig.~\ref{fig:Couples}(b)~\footnote{This is how our study differs from \cite{Panchenko_2016} -- it does not consider internal rotation or torque. They have considered self-propulsion only. Also, they have used dissipative particle dynamics, as opposed to the Langevin dynamics considered here.}.
Because of internal rotations, we can define the quantities internal angular momentum density $\rho \textbf{M}(\mathbf{x}, t)$, moment of inertia density $\textbf{I}(\mathbf{x}, t)$, and local angular velocity $\boldsymbol{\omega}(\mathbf{x}, t)$ 
\begin{equation}
\label{eq:Internal}
\rho \, \mathbf{M} = \rho\mathbf{L} - \rho \mathbf{x} \times \mathbf{v}, \quad \mathbf{I} = \sum_{i \alpha} \mathbf{I}_i^\alpha \Delta_i^\alpha, \quad \mathbf{I} \, \boldsymbol{\omega} = \rho \mathbf{M},
\end{equation}
where $\rho \mathbf{x} \times \mathbf{v}$ is the moment of momentum of the continuum point $\mathbf{x}$, $\mathbf{I}^{\alpha}_i =   m^{\alpha}_i (\mathbf{x}^{\alpha}_i-\mathbf{x})^2 \mathcal{I} - m^{\alpha}_i (\mathbf{x}^{\alpha}_i -\mathbf{x}) \otimes (\mathbf{x}^{\alpha}_i-\mathbf{x})$, and $\mathcal{I}$ is the identity matrix. 
Note that the local angular velocity in Eq.~\eqref{eq:Internal} depends on the coordinate system because of the definition of moment of inertia. 
Therefore, all the results that include $\boldsymbol{\omega}$ should be interpreted accordingly. 
The balance equation for the internal angular momentum can be derived to be~\cite{Dahler_1959, Dahler_1961, Dahler_1963}
\begin{equation}
\label{eq:IntAngMomBal}
\rho \dot{\bold{M}} = \boldsymbol{\nabla}\cdot \bold{C} + \rho \bold{G} +\bold{A},
\end{equation}
which states that the internal angular momentum density is driven by the local surface and volume couples, and the antisymmetric portion of the stress tensor. 
This is Cauchy's second law of motion, a generalization of Euler's equation for rigid body rotation to deformable bodies~\cite{Stokes_1984}.
The relevance of internal angular momentum density for material properties was first pointed out by Cosserats and then revived by H. Grad, and Dahler and Schriven, among others~\cite{Dahler_1961, Stokes_1984}. 
Our work shows that active suspensions provide a timely illustration of these ideas.


The kinetic energy density of the system at the hydrodynamic level is composed of the translational part $\rho \bold{v}^2 / 2$ and the rotational part $(\bold{I} \boldsymbol{\omega}) \cdot \boldsymbol{\omega} / 2$. 
Accordingly, we define the internal energy density $\rho\epsilon$ by
\begin{subequations}
\label{eq:EngDec}
\begin{eqnarray}
\rho\epsilon & = & \rho e - \frac{1}{2} \rho \bold{v}^2 - \frac{1}{2} (\bold{I} \boldsymbol{\omega}) \cdot \boldsymbol{\omega} \\
& = & \frac{1}{2} \sum_{i \alpha} \big[ m^{\alpha}_i(\bold{v}^{\alpha}_i-\hat{\bold{v}}^{\alpha}_i)^2+ \sum_{j \beta} u_{ij}(\bold{x}^{\alpha}_i,\bold{x}^{\beta}_j) \big] \Delta_i^\alpha \\
& \equiv & \sum_{i \alpha} \epsilon_i^\alpha \Delta_i^\alpha,
\end{eqnarray}
\end{subequations}
where $\rho e$ is the total energy density defined in Eq.~\eqref{eq:Density}, and $\hat{\mathbf{v}}_i^\alpha = \mathbf{v} + \boldsymbol{\omega} \times (\mathbf{x}_i^\alpha - \mathbf{x})$ is the rigid body like convective velocity of any particle $(i, \alpha)$. 
The internal energy per particle $\epsilon_i^\alpha$ is therefore composed of inter-particle potential energies and the ``thermal" energy of particles coming from the fluctuations in velocity with respect to the locally comoving and corotating frame, $\mathbf{v}_i^\alpha - \hat{\mathbf{v}}_i^\alpha$. 
Note that in a simple (nonpolar) fluid the thermal energy is defined only with respect to the comoving frame because the internal rotations are absent. 
To write down the balance equation for both total and internal energy density, we need the balance equation for moment of inertia density, which can be derived to be of the form
\begin{equation} 
\label{eq:InnBal}
\dot{\bold{I}} + \bold{I} \left(\boldsymbol{\nabla} \cdot \bold{v}\right) = \boldsymbol{\nabla} \cdot \sum_{i \alpha} \bold{I}^{\alpha}_i \otimes (\bold{v}-\bold{v}^{\alpha}_i) \Delta_i^\alpha \equiv \boldsymbol{\nabla} \cdot \mathbf{Y},
\end{equation}
where $\mathbf{Y}$ (defined in the second relation) is a third order tensor that corresponds to the flux of moment of inertia. 
The balance of moment of inertia exists simply due to the exchange of ADPs through the neighborhood of the macroscopic point $\mathbf{x}$.

Using Eq.~\eqref{eq:InnBal} and the definitions of the stress and couple stress tensors as well as the body forces and torques in Table~\ref{tab:Exact}, the balance equation for total energy can be derived to be
\begin{eqnarray}
\label{eq:EngBal}
 \rho \dot{e} & = & - \boldsymbol{\nabla} \cdot \bold{J}_\text{q} + \boldsymbol{\nabla} \cdot (\bold{T}^T \bold{v}) + \boldsymbol{\nabla} \cdot ( \bold{C}^T \boldsymbol{\omega})+\rho (\bold{b} \cdot \bold{v}+ \bold{G}\cdot \boldsymbol{\omega}) \nonumber \\
 & & \hspace{0.5in} - \frac{1}{2} \boldsymbol{\nabla} \cdot \left( \bold{Y}:  \boldsymbol{\omega} \otimes \boldsymbol{\omega}  \right)+ \Lambda.
\end{eqnarray} 
The term $\textbf{J}_\text{q} =  \bold{J}^{\text{K}}_\text{q} +  \bold{J}^{\text{P}}_\text{q}+  \bold{J}^{\text{A}}_\text{q}$~\footnote{In the accompanying paper we have $\mathbf{J}^\text{P}_\text{q} = \mathbf{J}^\text{V}_\text{q} + \mathbf{J}^\text{S}_\text{q}$.} is the heat flux vector, stemming from the kinetic transport of energy ($\bold{J}^{\text{K}}_\text{q}$) and the rate of work done by potential and active forces ($\bold{J}^{\text{P}}_\text{q}$ and $\bold{J}^{\text{A}}_\text{q}$, respectively): 
\begin{subequations}
\label{eq:Heat}
\begin{eqnarray}
\hspace{-0.8cm}\bold{J}^{\text{K}}_\text{q} & = & \sum_{i \alpha}(\epsilon_i^\alpha - m^\alpha_i \mathbf{v} \cdot [\omega \times (\mathbf{x}^\alpha_i - \mathbf{x})]) (\bold{v}^\alpha_i - \mathbf{v})  \Delta_i^\alpha \\
\hspace*{-0.8cm}\bold{J}^\text{P}_\text{q} & = & \frac{1}{2} \sum_{i j \alpha \beta} \big[ \bold{F}^{\alpha \beta}_{ij} \cdot \left( \bold{v}^\alpha_i - \hat{\bold{v}}_i^\alpha \right) \big] \bold{x}^{\alpha \beta}_{ij} b^{\alpha \beta}_{ij} \\
\hspace*{-0.8cm}\bold{J}^{\text{A}}_\text{q} & = & \sum_i \left[ \left( \bold{v}^2_i  - \hat{\bold{v}}_i^2 \right) \cdot \bold{f}_i^1 \right] \bold{x}^{12}_{ii} b^{12}_{ii}.
\end{eqnarray}
\end{subequations}
The second and third terms in Eq.~\eqref{eq:EngBal} are the rates of work done by the stress and couple stress tensors, respectively. 
Similarly, the fourth and fifth terms, respectively, are the rates of work done by the body forces and couples. 
The sixth term (containing $\textbf{Y}$) is kinematic in nature, arising out of the diffusive transport of moment of inertia. 
The last term, $\Lambda$ (given in the following), is a \emph{source} of heat by body forces and appears as a result of the ``thermal" fluctuations of the velocities, the thermal forces $\mathbf{f}_{i, \text{Th}}^\alpha$, and the work done by the active torques on the ADP as a whole. 
The term $\Lambda$ may be understood as an extension of the concept of heat in the stochastic energetics framework to flowing  extended systems. 
We have $\Lambda=\sum_{i \alpha} \left[ \left( \bold{v}^\alpha_i  - \hat{\bold{v}}^\alpha_i \right) \cdot \mathbf{f}_{i, \text{Th}}^\alpha + \left( \hat{\bold{v}}^1_i - \hat{\bold{v}}^2_i \right) \cdot \bold{f}_i^1 \right] \Delta^\alpha_i$.

Finally, the balance equation for internal energy can be derived as 
 \begin{eqnarray}
\label{eq:IntEngBal}
\rho \dot{\epsilon} & = & -\boldsymbol{\nabla} \cdot \bold{J}_\text{q} + \bold{T} : \nabla \bold{v} + \bold{C} : \nabla \boldsymbol{\omega} - \bold{A} \cdot \boldsymbol{\omega} \nonumber \\
& & -  \frac{1}{2} \bold{Y} :\nabla (\boldsymbol{\omega} \otimes \boldsymbol{\omega})+ \Lambda,
\end{eqnarray}
where we have used the notations $(\nabla \mathbf{g})_{ij} = (\partial / \partial x_i) g_j$ for any vector $\mathbf{g}$ and $\mathbf{B}:\mathbf{D} = \sum_{i, j} B_{ij} D_{ji}$ for any two second rank tensors $\mathbf{A}$ and $\mathbf{B}$. It is interesting to note that the antisymmetric portion of the stress tensor $\mathbf{A}$ contributes to work done on the system by coupling with local angular velocity $\boldsymbol{\omega}$. Equation~\eqref{eq:IntEngBal} can be interpreted as the first law of thermodynamics where the first term on the right denotes heat flow and the rest denotes work. 
The equations in this Letter can be considered a development towards the stochastic thermodynamics of continuous media for active matter systems. 

\acknowledgements
The authors would like to thank Fr\'ed\'eric van Wijland, Steve Granick, Michael Hagan, David Limmer, Robert Jack, Michael R. DeWeese, and Panayiotis Papadopoulos for useful discussions.  
KKM acknowledges support from a National Institutes of Health Grant R01-GM110066. 
He is also supported by Director, Office of Science, Office of Basic Energy Sciences, Chemical Sciences Division, of the U.S. Department of Energy under contract No. DE-AC02-05CH11231. 
DM acknowledges support from the U. S. Army Research Laboratory and the U. S. Army Research Office under contract W911NF-13-1-0390. 
KK acknowledges support from an NSF Graduate Research Fellowship.


%

\end{document}